\documentclass[conference,a4paper,table]{APSIPA2021}
\usepackage{multirow}
\usepackage{lipsum}
\usepackage{graphicx,xcolor}
\usepackage{url}
\usepackage{hyperref}

\begin{document}

\title{Improving Security in McAdams Coefficient-Based Speaker Anonymization by Watermarking Method}

\author{%
\authorblockN{%
Candy Olivia Mawalim and
Masashi Unoki
}
\authorblockA{%
Japan Advanced Institute of Science and Technology, \\
1-1 Asahidai, Nomi, Ishikawa 923--1292 Japan\\
Email: candyolivia@jaist.ac.jp, unoki@jaist.ac.jp}
}

\maketitle
\thispagestyle{empty}

\begin{abstract}
  Speaker anonymization aims to suppress speaker individuality to protect privacy in speech while preserving the other aspects, such as speech content. One effective solution for anonymization is to modify the McAdams coefficient. In this work, we propose a method to improve the security for speaker anonymization based on the McAdams coefficient by using a speech watermarking approach. The proposed method consists of two main processes: one for embedding and one for detection. In embedding process, two different McAdams coefficients represent binary bits ``0" and ``1". The watermarked speech is then obtained by frame-by-frame bit inverse switching. Subsequently, the detection process is carried out by a power spectrum comparison. We conducted objective evaluations with reference to the VoicePrivacy 2020 Challenge (VP2020) and of the speech watermarking with reference to the Information Hiding Challenge (IHC) and found that our method could satisfy the blind detection, inaudibility, and robustness requirements in watermarking. It also significantly improved the anonymization performance in comparison to the secondary baseline system in VP2020.
\end{abstract}

\section{Introduction}
Speech technology has been developed rapidly in recent years, with applications such as a virtual assistant that receives speech commands as its input. On the one hand, this technology helps us to live more conveniently. On the other hand, it can cause privacy issues because speech encapsulates personal identifiable information (PII) such as age, gender, health, and emotional state \cite{tomashenko2020introducing}. This issue has been escalating due to the increasing availability of advanced speech synthesis technology. At this stage, with only a few speech samples spoken by a target speaker, we could easily clone a new utterance spoken by the speaker \cite{Sercan2018}.

To preserve speech privacy, speaker anonymization or de-identification is used as a preprocessing method before the speech is publicly distributed. Research on speaker anonymization constitutes a relatively new area that has emerged following the efforts to suppress speaker identity to fool speaker identification systems. In prior studies, the Kaldi phone has been used for voice transformation \cite{jin2008voice, jin2009speaker} and could successfully fool the earlier speaker identification system based on a Gaussian mixture model (GMM). In addition, another approach based on cepstral frequency warping has been used for speech transformation for the purpose of de-identification\,\cite{magarinos2017reversible}. 

Although a few solutions have been proposed for anonymization, there is no formal definition or task in place yet, which causes ambiguity in determining the level of anonymization performance. Hence, the VoicePrivacy 2020 Challenge (VP2020) was initiated with the goal of fostering the development of speaker anonymization techniques by defining the task, metrics, scenarios, and a benchmark for the initial solutions (baseline). In this challenge, an anonymization system suppressed PII while maintaining the intelligibility and naturalness of a speech signal \cite{tomashenko2020introducing}.Two baseline systems were provided: (1) a primary baseline based on a state-of-the-art x-vector embedding and neural waveform modelling \cite{Fang2019}, and (2) a secondary baseline based on the manipulation of pole locations derived by linear predictive coding (LPC) using the McAdams coefficient \cite{patino2020speaker}. 

As a result of the VP2020, several systems have been proposed for speaker anonymization \cite{VP2020_results}. Most of these systems focused on improving the primary baseline \cite{MawalimGKU20, Turner2020, Espinoza2020}, since it achieved superior results in comparison to the secondary baseline. In this paper, we focus on improving the security of the secondary baseline, which is much less complex than the primary baseline and requires no training data. Our proposed method is based on a speech watermarking approach. With speech watermarking, it is possible to embed a secret message (also called a watermark) \cite{Hua2016} that can be used to verify the authentication of the speech. We speculate that by embedding a watermark in anonymized speech, we should be able to identify the originality of the speech or use it to prevent speech spoofing.

In Section 2 of this paper, we provide a brief overview of related studies. Section 3 introduces our proposed method. Our experimental setting, evaluations, and results are reported in Section 4. We conclude in Section 5 with a brief summary and mention of future work.

\section{McAdams coefficient-based Speaker Anonymization}

As one solution for preserving the privacy of speech data, VP2020 \cite{tomashenko2020introducing} was initiated with the goal of fostering the development of speaker anonymization techniques. Such techniques are ideally expected to suppress the leakage of PII while maintaining the linguistic information of the speech signal. In this challenge, four requirements were determined for the speaker anonymization technique: (1) the output had to be a speech waveform, (2) it must maximize the suppression of speaker individuality information, (3) it must preserve speech naturalness and intelligibility, and (4) it must ensure the distinction of voices of different speakers. This challenge also provided two different baseline systems\footnote{\url{https://github.com/Voice-Privacy-Challenge/Voice-Privacy-Challenge-2020}}, datasets, and protocols for evaluating the speaker anonymization performance.

\begin{figure}[t]
  \centering
  \includegraphics[width=.9\linewidth]{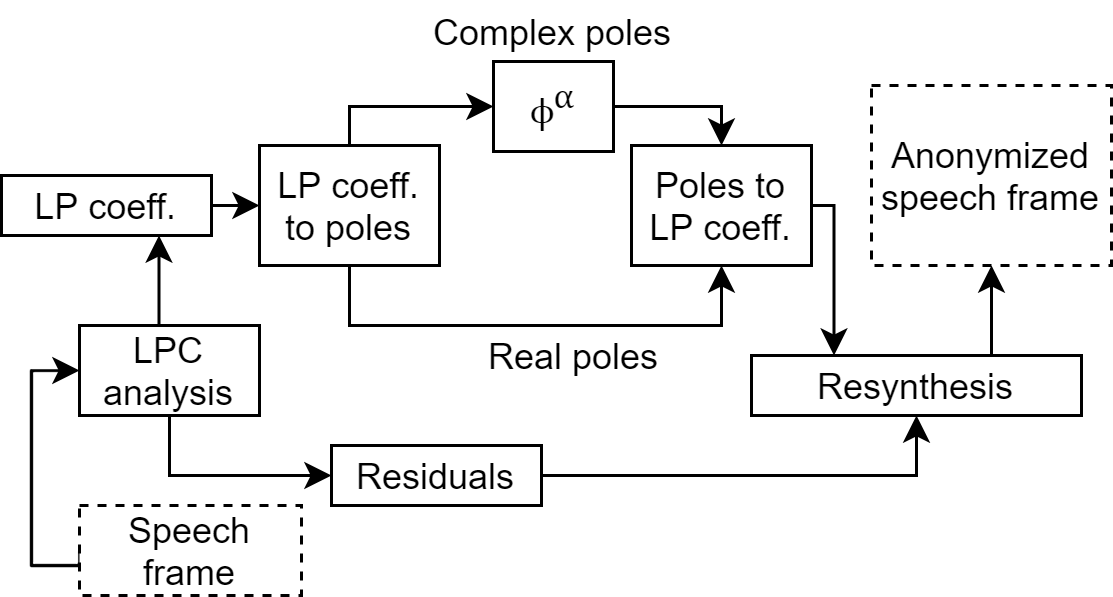}
  \caption{Block diagram of speaker anonymization based on McAdams coefficient \cite{patino2020speaker}. ``LP coeff.'' is referred to as linear prediction coefficients. ``LPC'' is referred to as linear predictive coding.  ``$\phi$'' is the angle of poles with a non-zero imaginary part. ``$\alpha$'' is the McAdams coefficient.}
  \label{fig:mcadams_anon}
\end{figure}

The primary baseline system was developed on the basis of x-vector embeddings and a neural source filter-based waveform model \cite{Fang2019}. The idea behind this system is to modify the speech identity information (x-vector) after disentangling it from the speech content (phoneme posteriorgram and the fundamental frequency). The secondary baseline system (shown in Fig. \ref{fig:mcadams_anon}) was developed on the basis of modifications to the McAdams coefficient \cite{patino2020speaker}. The McAdams coefficient is related to the adjustment of harmonic frequency distributions, which affects the perception of timbre \cite{McAdams84}. Although the results of the secondary baseline were not as good as the first baseline, it requires no training data and is much less complex. 

\begin{figure}[t]
  \centering
  \includegraphics[width=\linewidth]{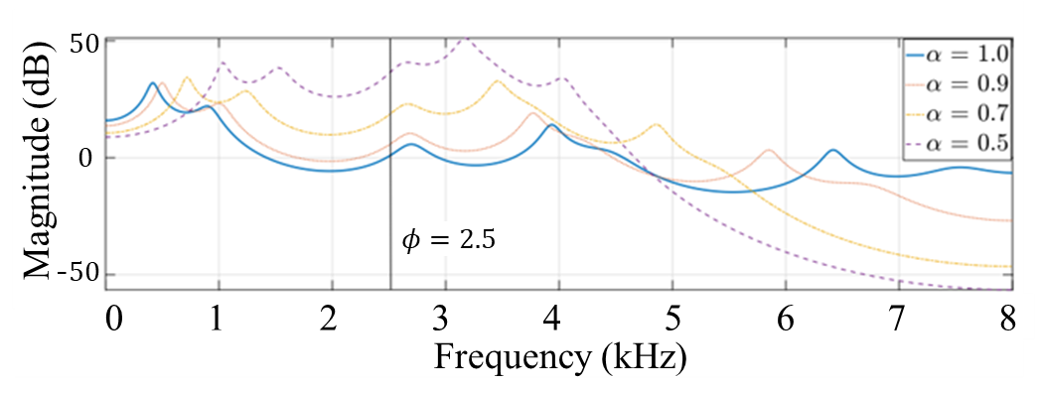}
  \caption{Spectral envelopes of speech frame with $\alpha\in\{1.0, 0.9, 0.7, 0.5\}$ \cite{patino2020speaker}.}
  \label{fig:freq_response}
\end{figure}

The McAdams coefficient proposed in \cite{McAdams84} is a parameter derived on the basis of the additive synthesis method in music signal processing \cite{Dodge1986ComputerMS}. This method is applied to timbre generation by resynthesizing multiple harmonic cosinusoidal oscillations, as
\begin{equation}\label{eq:McAdams_eq}
    y_{syn}(t) = \sum_{h=1}^{H}r_h(t)\cos(2\pi (hf_0)^\alpha t + \Phi _h),
\end{equation}
where $y_{syn}(t)$ is the synthesized signal by combining harmonic consinusoidal oscillations with inverse Fourier series, $h$ is the harmonic index, $r_h(t)$ is the amplitude, $\Phi_h$ is the phase, and $\alpha$ is the McAdams coefficient \cite{McAdams84}. Prior work on speaker anonymization \cite{patino2020speaker}, has shown that the McAdams coefficient can transform the spectral envelope of speech signals and affect timbre perception. To examine this in more detail, we show the spectral envelopes of the frequency response by various McAdams coefficients in Fig. \ref{fig:freq_response}. We can see that the farther away the McAdams coefficient is from the original speech ($\alpha = 1$), the greater the shift of spectral envelope we can obtain. The degree of this shifting affects our perception in perceiving the speech formants \cite{McAdams84}.

\section{Proposed Method}

In this study, we propose a method based on the secondary baseline to enhance the security of anonymized speech by using a watermarking approach. Speech watermarking aims to protect the security in a speech signal by imperceptibly embedding within it a particular message, such as a signature that indicates the speech's ownership. Speech watermarking should fulfill at least three requirements: inaudibility (not perceivable by the human auditory system), blindness (detection without the availability of original signal), and robustness against common signal processing operations. The trade-off between inaudibility and robustness has been the most pressing issue in existing speech watermarking techniques \cite{Hua2016}. In the present work, we investigate the effectiveness of using a watermarking approach for a McAdams coefficient-based speech anonymization method with regard to these requirements.

\begin{figure}[t]
  \centering
  \includegraphics[width=\linewidth]{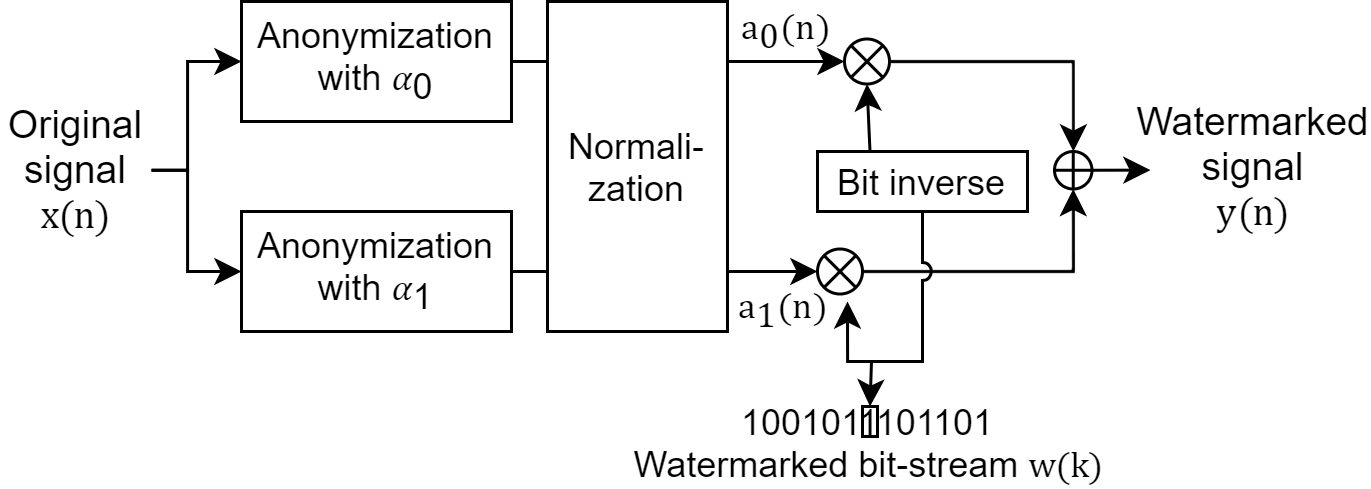}
  \caption{Block diagram of embedding process. $\alpha_0$ and $\alpha_1$ are the McAdams coefficients for representing binary bit ``0'' and ``1''. $a_0(n)$ and $a_1(n)$ are the output anonymized speech in time domain.}
  \label{fig:embedding}
\end{figure}

Generally, speech watermarking consists of two main processes: embedding and detection. Figure \ref{fig:embedding} shows the block diagram of our embedding process. As the first step, we generated the anonymized signals from the original signal ($x(n)$) using two different McAdams coefficients ($\alpha_0$ and $\alpha_1$). The anonymization procedure follows the steps in Fig. \ref{fig:mcadams_anon}.

Firstly, the original speech was segmented into speech frames with frame length depending on the watermarking payload. Subsequently, the speech frame was analyzed using linear predictive coding (LPC) with an order of 20 ($M=20$). The mathematical form of the LPC is characterized by the following differential equation:

\begin{equation} \label{eq:LPC}
s(n) = \sum_{i=1}^{M}c(i)s(n-i) + e(n)
\end{equation}

\noindent where $s(n)$ is a speech signal, $c(i)$ corresponds to the filter coefficient in $i$-th order, $M$ is the maximum order of the prediction (in this study $M=20$), and $e(n)$ is the prediction error. The corresponding transfer function ($H(z)$) for Eq. (\ref{eq:LPC}) is represented by twentieth-order all-pole autoregressive filters, which is given by:
\begin{equation}
    H(z) = \frac{1}{1-\sum_{i=1}^{M}c(i)z^{-i}}
\end{equation} 

From LPC analysis, we obtained the linear prediction coefficients (LP coefficients) and residuals. These LP coefficients ($c(i)$) were then used to derive the pole positions. The derived poles were comprised of complex poles (poles with non-zero-valued imaginary terms) and real poles (poles with zero-valued imaginary terms). The shift of complex poles position ($\phi^\alpha$) was resulting in the angle shifting to either clockwise or counter-clockwise of the complex positions \cite{patino2020speaker}. The effect of this angle shifting in the spectral envelope is shown in Fig. \ref{fig:freq_response}. After the modification of McAdams coefficients $\alpha$, the modified complex poles and the real poles were converted back to LP coefficients. The anonymized speech frame was resynthesized from these LP coefficients and the original residuals.

After obtaining the anonymized signals, we normalized them to be in similar relative loudness (fixed a target peak level in decibel relative to full scale (dBFS)) and range of frequency components (using a bandpass filter (BPF)). The cut-off frequencies for the BPF were 125 Hz and 4 kHz. Finally, we constructed the watermarked signal ($y(n)$) by frame-by-frame concatenation of the anonymized signals obtained by bit inverse according to watermarked bit-stream.

\begin{figure}[t]
  \centering
  \includegraphics[width=.9\linewidth]{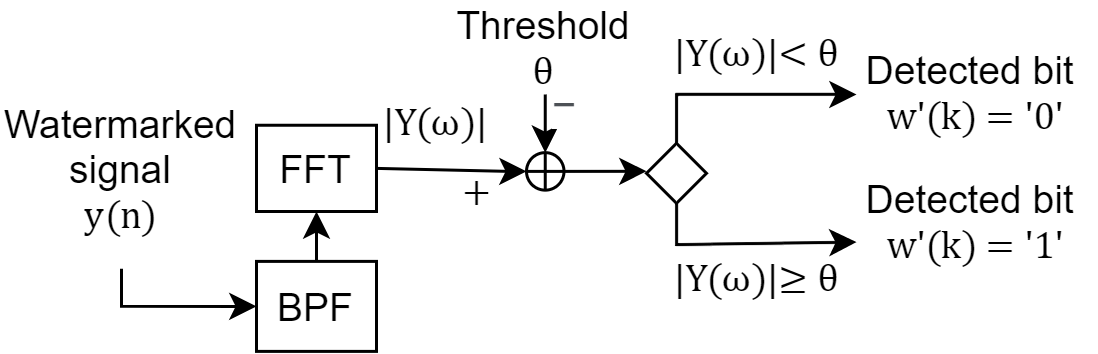}
  \caption{Block diagram of blind detection process. ``BPF'' stands for the band-pass filtering. ``FFT'' stands for fast Fourier transform. $|Y(\omega)|$ is the power spectrum of the watermarked signal $y(n)$ obtained by FFT. $\theta$ is the power spectrum threshold for blind detection process. $w'(k)$ is the detected watermark bit of the $k$-th frame.}
  \label{fig:detection}
\end{figure}

We found that the anonymized signals from different McAdams coefficients carried different amounts of power spectrum, specifically in the lower frequency components. Using a higher McAdams coefficient results in a higher power spectrum in the lower frequency. On the basis of this characteristic, we determined a power threshold for the blind detection process (shown in Fig. \ref{fig:detection}). The detection process was conducted by comparing the power spectrum obtained by fast Fourier transform (FFT) of the watermarked signal ($|Y(\omega)|$) in a specific frequency range with the designated threshold $\theta$. 

\section{Experiments}

This section describes the experimental setting in our study and reports the objective evaluation results.

\subsection{Experimental Setting}

We used LibriSpeech \cite{libriSpeech} and VCTK \cite{VCTK} datasets (development and testing sets) that were provided in VP2020. LibriSpeech is an English speech corpus designed for automatic speech recognition (ASR) research sampled at 16 kHz \cite{libriSpeech}. VCTK is an English speech corpus that contains 109 native speakers with various accents and was designed for text-to-speech (TTS) research sampled at 48 kHz. The development and training data of both datasets in VP2020 consists of more than 20,000 utterances from almost 200 speakers. The sampling rate of the speech data is set to 16 kHz. Similar to the secondary baseline \cite{patino2020speaker}, we do not need any training data for our proposed method. The McAdams coefficient used to represent bit ``0" was 0.6 ($\alpha_0=0.6$) and bit ``1" ($\alpha_1=0.8$) was 0.8. We evaluated the speaker verifiability of our proposed method by using a pretrained automatic speaker verification system (ASVeval) and the intelligibility by using a pretrained automatic speech recognition system (ASReval), similar to the protocol in VP2020. The suggested requirement for a reasonable payload of an audio watermarking is 72 bits per 30 seconds (around 2--3 bps) \cite{STEP2001}. In this study, we used a 4-bps watermarked signal to evaluate the anonymization performance.

\begin{figure}[t]
  \centering
  \includegraphics[width=\linewidth]{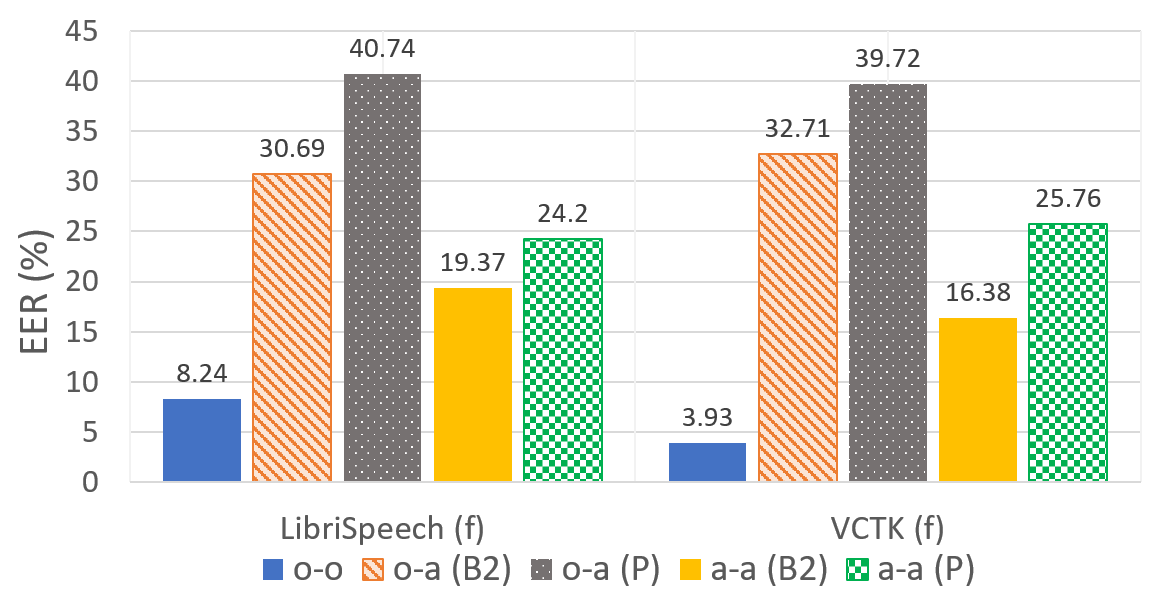} \\
  \small (a) Female utterances
  \label{fig:EER_f}
  \includegraphics[width=\linewidth]{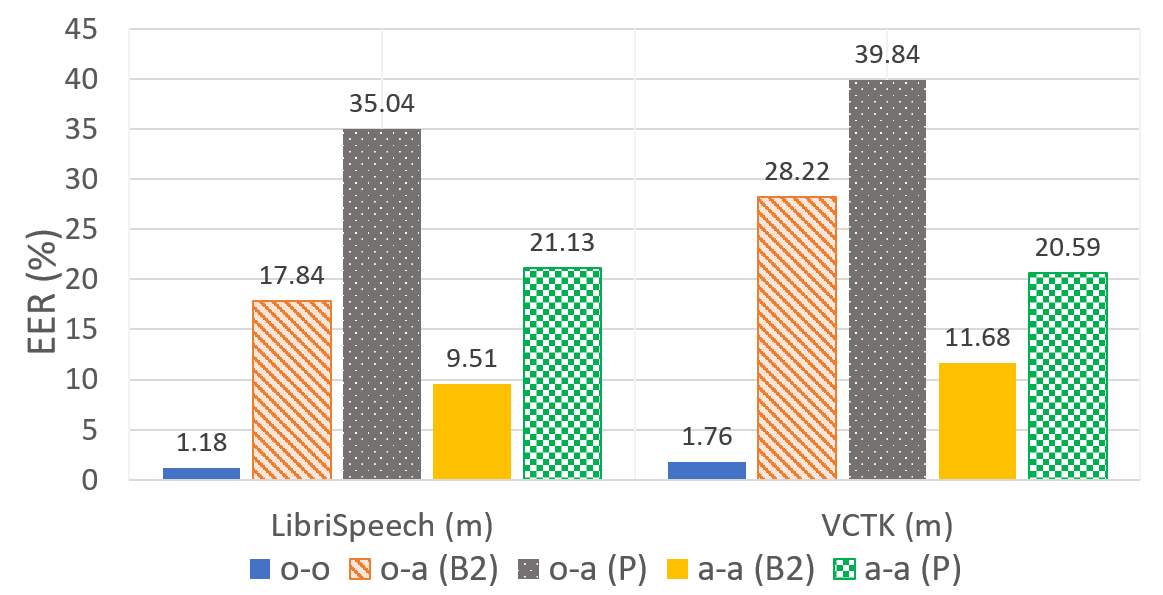} \\
  \small (b) Male utterances
  \label{fig:EER_m}
  \caption{Speaker verifiability evaluation in equal-error rate (EER). ``f'' stands for female utterances, whereas ``m'' stands for male utterances. There are three scenarios considered, i.e., (1) o-o (both original data for enrollment and trial), (2) o-a (original data for enrollment and anonymized data for trial), and (3) a-a (both anonymized data for enrollment and trial). }
  \label{fig:EER}
\end{figure}

For speech watermarking, we also evaluated the speech quality and robustness of our method with a total 100 randomly selected utterances from the LibriSpeech and VCTK datasets. Since the original signal was not available, we used MOSNet, the pretrained mean opinion score (MOS) predictor proposed in \cite{MOSNet}. MOSNet is an objective evaluation tool based on deep learning approach for predicting human MOS ratings in a voice conversion system. Subsequently, we evaluated the robustness of our propsed method as suggested in \cite{IHC} by calculating the bit error rate (BER) and balanced F1-score during normal (no attack) operations along with several signal processing operations, including noise addition, resampling, requantization, compression, and speech codecs. We also examined the security level by calculating the false acceptance rate (FAR) and false rejection rate (FRR). The maximum acceptable BER threshold as the robustness indication is $10\%$ \cite{IHC}. We embedded random binary streams with payloads of 2, 4, 8, 16, and 32 bps and varied the detection threshold in the order of lower to higher payloads (0.15, 0.09, 0.05, 0.02, and 0.01, respectively). Due to the space limitation, the results reported in this paper are mainly in mean value.

\subsection{Results}

Our intent with this evaluation is mainly to investigate the effectiveness and reliability of the proposed method in anonymizing the PII of the speaker and watermarking the speech. For the speaker anonymization, we conducted our evaluation using ASVeval to check the speaker verifiability performance in three cases: original enrolls and original trials case (o-o), original enrolls and anonymized trials case (o-a), and anonymized enrolls and anonymized trials case (a-a). The results are shown in Fig. \ref{fig:EER}, where the top and bottom graphs shows the results of female and male utterances, respectively. As we can see, the proposed method (P) improved the speaker similarity in comparison to the secondary baseline in VP2020 (B2): the EER improved significantly to more than 35\% in the o-a case and to more than 20\% in the a-a case for both female and male utterances.

\begin{table}[t]
    \centering
    {\small
    \caption{MOSNet evaluation results.}
    \label{tab:MOSNet_results}
    \begin{tabular}{|c|c|c|}
    \hline
    \rowcolor[HTML]{EFEFEF} 
    \multicolumn{1}{|l|}{\cellcolor[HTML]{EFEFEF}} & \textbf{payload (bps)} & \textbf{MOS} \\ \hline
    \textbf{original} & - & 3.15 ± 0.49 \\ \hline
    \textbf{anonymized} & - & 2.70 ± 0.18 \\ \hline
     & 2 & 2.73 ± 0.20 \\ \cline{2-3} 
     & 4 & 2.73 ± 0.21 \\ \cline{2-3} 
     & 8 & 2.70 ± 0.19 \\ \cline{2-3} 
     & 16 & 2.67 ± 0.18 \\ \cline{2-3} 
    \multirow{-5}{*}{\textbf{\begin{tabular}[c]{@{}c@{}}proposed \\ method\end{tabular}}} & 32 & 2.60 ± 0.18 \\ \hline
    \end{tabular}}
\end{table}

\begin{figure}[t]
  \centering
  \includegraphics[width=\linewidth]{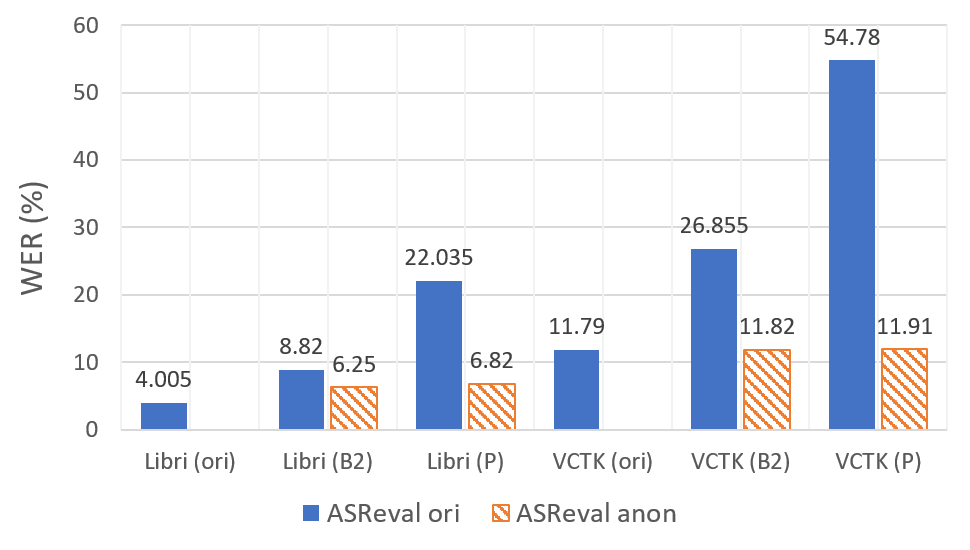}
  \caption{Speech intelligibility evaluation in word error rate (WER). ``ASReval ori” and ``ASReval anon” denote ASR trained on original and anonymized data, respectively. ``Libri'' is referred to as LibriSpeech dataset. ``VCTK'' is referred to as VCTK dataset. ``ori'', ``B2'', ``P'' are original speech, the output of the secondary baseline in VP2020, the output of our proposed method, respectively.}
  \label{fig:WER}
\end{figure}

\begin{figure*}[t]
  \centering
  \includegraphics[width=0.87\linewidth]{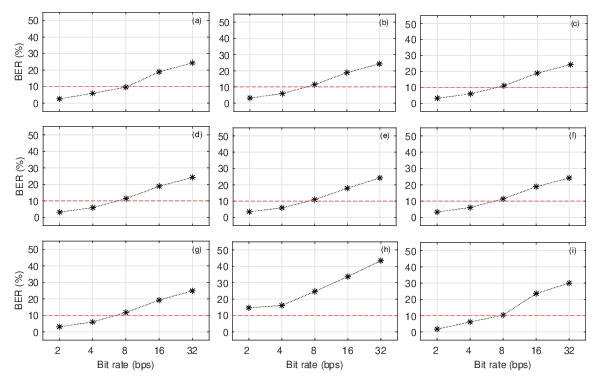}
  \caption{Robustness test results in terms of BER (bit error rate) in nine cases: (a) normal, (b) AWGN, (c) resample-8, (d) resample-24, (e) requant-8, (f) requant-24, (g) mp3, (h) flv, and (i) G723.1.}
  \label{fig:BER}
\end{figure*}

\begin{figure*}[t]
  \centering
  \includegraphics[width=0.87\linewidth]{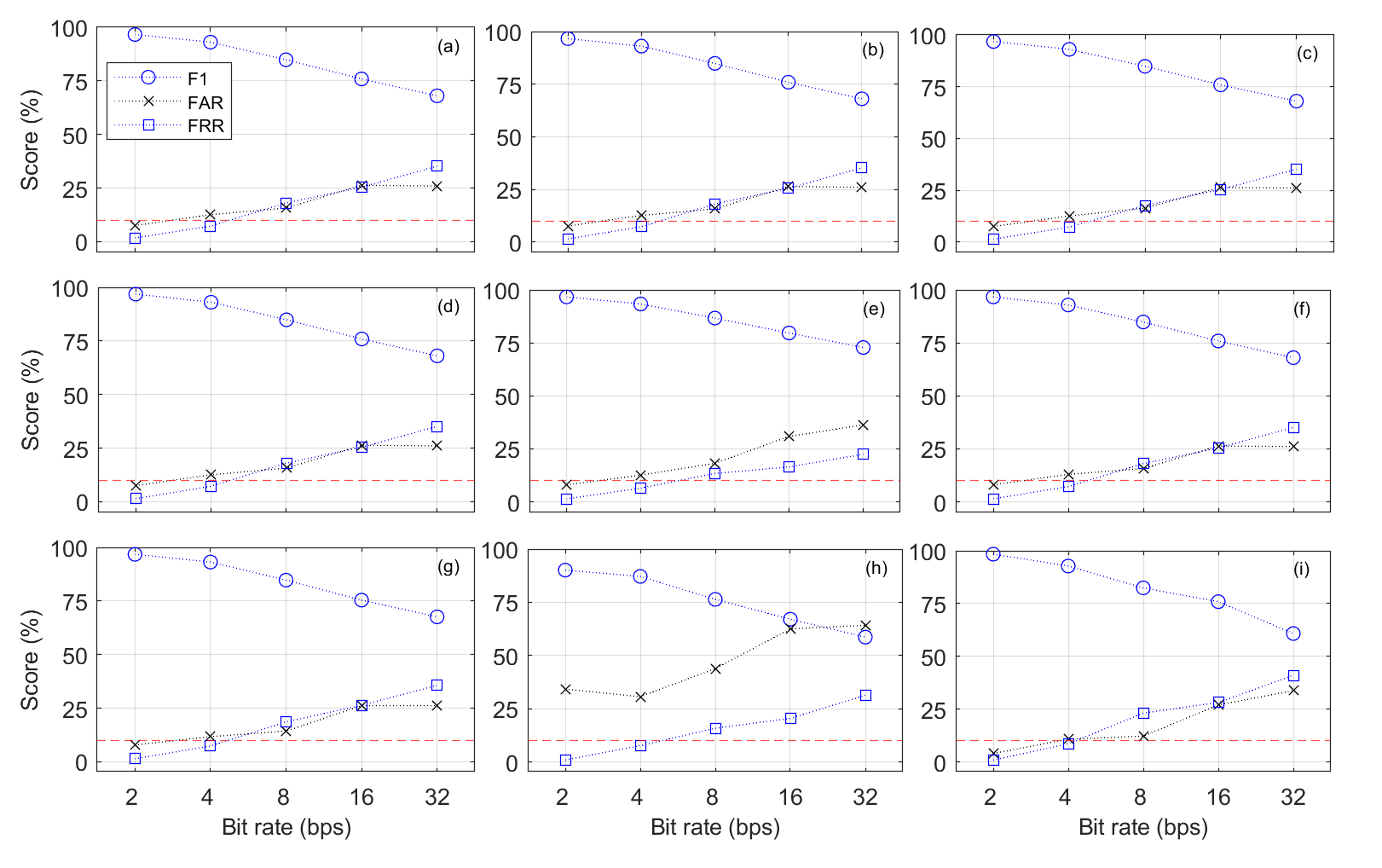}
  \caption{Robustness test results in nine cases: (a) normal, (b) AWGN, (c) resample-8, (d) resample-24, (e) requant-8, (f) requant-24, (g) mp3, (h) flv, and (i) G723.1. For metrics were used for the robusness evaluation, including F1 (F1-score), FAR (false acceptance rate), and FRR (false rejection rate).}
  \label{fig:results}
\end{figure*}

For the speech intelligibility, we conducted an objective evaluation by using ASReval trained on original data and anonymized data. The results are shown in Fig. \ref{fig:WER}, where the blue and orange bars indicate the results obtained from ASReval trained on original and anonymized data, respectively. In contrast to the speech verifiability, our proposed method caused a noticeable degradation to intelligibility, especially for the VCTK dataset (the degradation is increased from 11.79\% to 54.78\%). This degradation was mainly occurred due to the anonymization method as we can obtain almost similar average WER with the stochastic approach proposed in \cite{patino2020speaker}. The further shift to McAdams coefficients, the more WER occurred. To improve the intelligibility, we could retrain the ASReval using the anonymized data as in \cite{patino2020speaker}. For example, the WER is shown by the blue bar for ``VCTK (P)" could be reduced to 11.91\%.

Inaudibility and robustness are essential requirements in speech watermarking. To evaluate the inaudibility of the proposed method, we used MOSNet \cite{MOSNet} to derive the MOS of the watermarked signal. Table \ref{tab:MOSNet_results} shows the MOSNet results of original signal, anonymized signal with McAdams coefficients $\alpha = 0.8$, and the output signal of our proposed method with various payloads. We can see that there was a speech quality degradation (MOS degraded from 3.15 to 2.70) caused by the McAdams coefficient-based anonymization method, while in contrast, the proposed method could maintain a similar MOS even with a relatively high payload.

\begin{figure}[t]
  \centering
  \includegraphics[width=.93\linewidth]{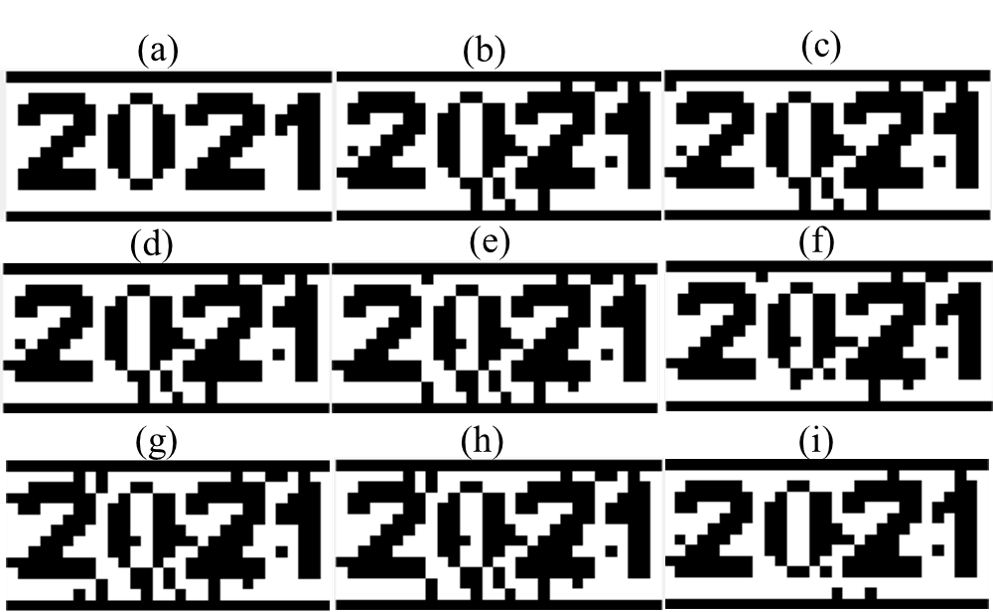}
  \caption{Detected watermarks from image embedding in 4-bps watermarked signal after several operations: (a) original, (b) normal, (c) resample-8, (d) resample-24, (e) requant-8, (f) requant-24, (g) AWGN, (h) MP3, and (i) G723.1.}
  \label{fig:img_embed}
\end{figure}

We carried out a robustness test by calculating the detection accuracy from the output speech after several common signal processing operations. Figures \ref{fig:BER} and \ref{fig:results} show the robustness test results. We examined nine cases: no attack (normal), addition of white Gaussian noise (AWGN), downsampling to 8 kHz (resample-8), upsampling to 24 kHz (resample-24), bit compression to 8 bits (requantize-8), bit extension to 24 bits (requantize-24), MP3 compression (MP3), flash video format (flv), and G723.1 codec. For AWGN processing, the signal to noise ratio used is 40 decibel (dB). Meanwhile, the bitrate range for MP3 compression was from 220-260 kbps (kilo bits per second). The bitrate of G723.1 codec was 5.3 kbps with algebraic code-excited linear prediction (ACELP) algorithm. As we can see, our proposed method was robust against other operations (the BER was similar to the normal case), except for the conversion to video codec (flv). The results here demonstrate that our method is suitable for watermarking purposes, since the BER for 4 bps satisfied the robustness criteria ($\mathrm{BER}<10\%$). The results also suggest that the security level indicated by the FAR, FRR, and F1-score is adequate for payloads up to 4 bps.

In summary, our proposed method could improve significantly the speaker verifiability by ASVeval. Although intelligibility was degraded, the WER could reduced by simply retraining the ASReval. The evaluation for speech watermarking also confirmed that our proposed method could resulted in inaudible and robust watermark. As an example, Figure \ref{fig:img_embed} shows the example of original image watermarked in the speech with the detected image after several operations. We could easily perceive the 2021 logo in the detected watermarks in almost all processing (BER is approximately 7\%). 

\section{Conclusion and Future Work}

In this paper, we proposed a technique to improve the security of McAdams coefficient-based speaker anonymization by using a watermarking approach. By adding a watermark into the anonymized signal, we could ensure the originality of the speech (one of the anti-spoofing countermeasures). The performance of the proposed method was evaluated objectively based on the standard in speaker anonymization (VP2020 protocol) and watermarking. The results showed that our method is suitable for watermarking, as it could satisfy the blind, inaudible, and robustness requirements. In addition, it significantly improved the speaker dissimilarity in ASVeval. The limitation of our method in comparison to the baseline system is that it caused degradation in speech intelligibility, although this could be resolved by retraining the ASReval with the anonymized speech.

In spite of these promising results, we acknowledge that the evaluation, particularly in consideration to human perception such as the speech quality test, might not be sufficient. As future work, we will explore more suitable ways to evaluate the watermarking technique for speaker anonymization. We also intend to optimize the proposed method, as its detection rate and payload are relatively limited.

\section{Acknowledgements}

This work was supported by a Grant-in-Aid for Scientific Research (B) (No. 17H01761), JSPS KAKENHI Grant (No.20J20580), Fund for the Promotion of Joint International Research (Fostering Joint International Research (B))(20KK0233), and KDDI foundation (Research Grant Program).

\bibliographystyle{IEEEtran}

\bibliography{mybib}

\end{document}